\documentclass[12pt] {article}
\usepackage{psfrag}
\usepackage{graphicx}
\usepackage{latexsym,amsfonts}
\usepackage{amssymb}
\begin{document}

\title{Weak decays \\ of H--like ${^{140}}{\rm Pr}^{58+}$ and He--like
  ${^{140}}{\rm Pr}^{57+}$ ions}

\author{A. N. Ivanov$^{a}$\,\thanks{E--mail:
    ivanov@kph.tuwien.ac.at, Tel.: +43--1--58801--14261, Fax:
    +43--1--58801--14299} , M. Faber${^a}$\,\thanks{E--mail: 
faber@kph.tuwien.ac.at, Tel.: +43--1--58801--14261, Fax:
    +43--1--58801--14299} , R. Reda${^c}$ , P. Kienle$^{b,c}$\,
\thanks{E--mail: Paul.Kienle@ph.tum.de}}

\date{\today}

\maketitle

\begin{center} {\it $^a$Atominstitut der \"Osterreichischen
    Universit\"aten, Technische Universit\"at Wien, Wiedner
    Hauptstrasse 8-10, A-1040 Wien, \"Osterreich, \\
    $^b$Physik Department, Technische Universit\"at M\"unchen,
    D--85748 Garching, Germany,\\ $^c$Stefan Meyer Institut f\"ur
    subatomare Physik, \"Osterreichische Akademie der Wissenschaften,
    Boltzmanngasse 3, A-1090, Wien, \"Osterreich}
\end{center}

\begin{center}
\begin{abstract}
 The nuclear K--shell electron--capture $(EC)$ and positron
   $(\beta^+)$ decay constants, $\lambda_{EC}$ and $\lambda_{\beta^+}$
   of H--like ${^{140}}{\rm Pr}^{58+}$ and He--like ${^{140}}{\rm
     Pr}^{57+}$ ions, measured recently in the ESR ion storage ring at
   GSI, were calculated using standard weak interaction theory. The
   calculated ratios $R = \lambda_{EC}/\lambda_{\beta^+}$ of the decay
   constants agree with the experimental values within an accuracy
   better than $3\,\%$. 
\end{abstract}
\end{center}

\newpage

\section{Introduction}

The neutral ${^{140}}{\rm Pr}^{0+}$ atoms decay with a $99.4\,\%$
branch to the ground state of ${^{140}}{\rm Ce}^{0+}$ via a pure $1^+
\to 0^+$ Gamow--Teller transition \cite{APR1}. The K--shell $EC$ to
$\beta^+$ ratio was measured as $R_{EC/\beta^+} = 0.74(3)$ by Biryukov
and Shimanskaya \cite{APR2} and $R_{EC/\beta^+} = 0.90(8)$ by Evans
{\it et al.} \cite{APR3}. Because of the discrepancy of these results
the experiment was repeated by Campbell {\it et al.}, who found
$R_{EC/\beta^+} = 0.73(3)$ and claimed a disagreement of about
$15\,\%$ with the theoretical value by Bambynek {\it et al.}
\cite{APR5}. This problem came up again, when Litvinov {\it et al.}
succeeded in measuring the $\beta^+$ and orbital electron--capture
decay rates in fully ionised, H--like ${^{140}}{\rm Pr}^{58+}$ and
He--like ${^{140}}{\rm Pr}^{57+}$ ions in the Experimental Storage
Ring (ESR) at GSI in Darmstadt \cite{GSI}. For the He--like
${^{140}}{\rm Pr}^{57+}$ ion a ratio $R_{EC/\beta^+} = 0.96(8)$ agrees
well with $R_{EC/\beta^+} = 0.90(8)$ by Evans {\it et al.}
\cite{APR3}, but deviates by about 2.5 standard deviations from the
values of \cite{APR2} and \cite{APR4}. In addition Litvinov {\it et
  al.}  \cite{GSI} found that the H--like ${^{140}}{\rm Pr}^{58+}$ ion
with one electron in the K--shell decays $1.49(8)$ times faster than
the He--like one with two electrons in the K--shell. In this work we
show that using standard weak interaction theory one can reproduce the
experimental values of the ratios of the $EC$ and $\beta^+$ decay
constants of the H--like ${^{140}}{\rm Pr}^{58+}$ and He--like
${^{140}}{\rm Pr}^{57+}$ ions with an accuracy better than $3\,\%$. 

The paper is organised as follows. In Section 2 we give the results of
the calculation of the weak decay constants of the H-like
${^{140}}{\rm Pr}^{58+}$ and the He--like ${^{140}}{\rm Pr}^{57+}$
ions and their ratios. We compare the theoretical results with the
experimental data. In the Conclusion we discuss the obtained results.
In Appendices A and B we adduce the detailed calculations of the weak
decay constants of the H-like ${^{140}}{\rm Pr}^{58+}$ and the
He--like ${^{140}}{\rm Pr}^{57+}$ ions.

\section{Weak decay constants}

For the calculation of the weak decay constants of H-like and He--like
ions we use the Hamiltonian of the weak interaction  taken in the
standard form \cite{HS66}
\begin{eqnarray}\label{label1}
  \hspace{-0.3in}{\cal H}_W(x) = \frac{G_F}{\sqrt{2}}\,V_{ud}\, 
  [\bar{\psi}_n(x)\gamma^{\mu}(1 -
  g_A\gamma^5)\,\psi_p(x)]\,[\bar{\psi}_{\nu_e}(x)
  \gamma_{\mu}(1 - \gamma^5)\psi_e(x)] + {\rm h.c.},
\end{eqnarray}
where $G_F = 1.166\times 10^{-11}\,{\rm MeV}^{-2}$ is the Fermi
constant, $V_{ud} = 0.9738 \pm 0005$ is the CKM matrix element, $g_A =
1.2695\pm 0.0058$ is the axial coupling constant \cite{PDG06},
$\psi_n(x)$, $\psi_p(x)$, $\psi_{\nu_e}(x)$ and $\psi_e(x)$ are
the operators of the neutron, proton, neutrino and electron/positron
fields, respectively.

The detailed calculations of the weak decay constants of the H--like
${^{140}}{\rm Pr}^{58+}$ and He--like ${^{140}}{\rm Pr}^{57+}$ ions
are given in Appendices A and B.  Since the $EC$--decay of the H--like
${^{140}}{\rm Pr}^{58+}$ ion from the hyperfine state ${^{140}}{\rm
Pr}^{58+}_{F = \frac{3}{2}}$ with $F = \frac{3}{2}$ is suppressed (see
Appendix B and \cite{GSI,HFS}), we take into account that the H--like
${^{140}}{\rm Pr}^{58+}$ ion decays from the hyperfine ground state
${^{140}}{\rm Pr}^{58+}_{F = \frac{1}{2}}$ with $F = \frac{1}{2}$.

Using the Hamiltonian of the weak interaction (\ref{label1}) and
following \cite{APR5} for the $EC$--decay constants of the H--like
${^{140}}{\rm Pr}^{58+}_{F = \frac{1}{2}}$ and the He--like
${^{140}}{\rm Pr}^{57+}_{I = 1}$ ions in the ground states we obtain
the following expressions (see Appendix A)
\begin{eqnarray}\label{label2}
 \hspace{-0.3in} &&\lambda^{\rm (H)}_{EC}=\frac{1}{2 F + 1}\,\frac{3}{2}
  |{\cal M}_{\rm GT}|^2 |\langle \psi^{(Z)}_{1s}\rangle|^2
  \frac{Q^2_{\rm H}}{\pi},\nonumber\\
 \hspace{-0.3in} &&\lambda^{\rm (He)}_{EC}=  \frac{1}{2I + 1}\frac{3}{2}|{\cal M}_{\rm GT}|^2 
  |\langle \psi^{\rm (Z-1)}_{1s}\psi^{\rm (Z)}_{(1s)^2}\rangle|^2\frac{Q^2_{\rm He}}{\pi},
\end{eqnarray}
where $F = 1/2$\,\footnote{In the ratio $\frac{3}{2}$ of the
$EC$--decay constant $\lambda^{(\rm H)}_{EC}$ the factors 3 and
$\frac{1}{2}$ are caused by the hyperfine structure of the ground
state of the H--like ${^{140}}{\rm Pr}^{58+}_{F = \frac{1}{2}}$ ion and
the phase volume of the final state of the decay, respectively
(see Appendix A).} and $I = 1$ are the total angular momenta of the
H--like ${^{140}}{\rm Pr}^{58+}$ and the He--like ${^{140}}{\rm
  Pr}^{57+}$ ions, respectively, $Q_{\rm H} = (3348\pm 6)\,{\rm keV}$
and $Q_{\rm He} = (3351\pm 6)\,{\rm keV}$ are the $Q$--values of the
decays ${^{140}}{\rm Pr}^{58+} \to {^{140}}{\rm Ce}^{58+} + \nu_e$ and
${^{140}}{\rm Pr}^{57+} \to {^{140}}{\rm Ce}^{57+} + \nu_e$, ${\cal
  M}_{\rm GT}$ is the nuclear matrix element of the Gamow--Teller
transition \cite{GT}, $ \psi^{(Z)}_{1s}$ and $\psi^{\rm (Z-1)}_{1s}$
are the wave functions of the H--like ions ${^{140}}{\rm Pr}^{58+}$
and ${^{140}}{\rm Ce}^{57+}$, respectively, $Z = 59$ is the electric
charge of the mother nucleus ${^{140}}{\rm Pr}^{59+}$ and $\psi^{\rm
  (Z)}_{(1s)^2}$ is the wave function of the He--like ion
${^{140}}{\rm Pr}^{57+}$, $\langle \psi^{(Z)}_{1s}\rangle$ and
$\langle \psi^{\rm (Z-1)}_{1s}\psi^{\rm (Z)}_{(1s)^2}\rangle$ are
defined by
\begin{eqnarray}\label{label3}
 \hspace{-0.3in} &&\langle
  \psi^{(Z)}_{1s}\rangle = \frac{\int\!\! d^3x\psi^{(Z)}_{1s}(\vec{r}\,)
\rho(\vec{r}\,)}{
\int\! d^3x\rho(\vec{r}\,)},\nonumber\\
\hspace{-0.3in} &&\langle \psi^{\rm (Z-1)}_{1s}\psi^{\rm
    (Z)}_{(1s)^2}\rangle = \frac{\int\!\! d^3x_1d^3x_2\psi^{(Z-1)}_{1s}(\vec{r}_1)\psi^{\rm
      (Z)}_{(1s)^2}(\vec{r}_1,\vec{r}_2)\rho(\vec{r}_2)}{\int\! d^3x_2\rho(\vec{r}_2)},
\end{eqnarray}
where $\rho(\vec{r}\,)$ is the nuclear density \cite{GT}.  Using the
Woods--Saxon shape for the nuclear density $\rho(r)$ and the Dirac
wave functions for the H--like ion and the He--like ion
\cite{IZ80,He1} one gets $\langle \psi^{(Z)}_{1s}\rangle = \langle
\psi^{\rm (Z-1)}_{1s}\psi^{\rm (Z)}_{(1s)^2}\rangle = 1.66/\sqrt{\pi
  a^3_B}$, where $a_B = 1/Z \alpha m_e = 897\,{\rm fm}$ is the Bohr
radius of the H--like ion ${^{140}}{\rm Pr}^{58+}$; $m_e = 0.511\,{\rm
  MeV}$ is electron mass and $\alpha = 1/137.036$ is the
fine--structure constant \cite{PDG06}.

The $\beta^+$--decay constants of the H--like ${^{140}}{\rm Pr}^{58+}$
and the He--like ${^{140}}{\rm Pr}^{57+}$ ions are defined by (see Appendix A)
\begin{eqnarray}\label{label4}
  \lambda^{(\rm H)}_{\beta^+} &=&  \frac{2}{2F + 1}\,
\frac{|{\cal M}_{\rm GT}|^2}{4\pi^3}\,
  f(Q^{\rm H}_{\beta^+},Z - 1),\nonumber\\
\lambda^{(\rm He)}_{\beta^+} &=&  \frac{3}{2 I + 1}\,
\frac{|{\cal M}_{\rm GT}|^2}{4\pi^3} \, 
  f(Q^{\rm He}_{\beta^+},Z-1).
\end{eqnarray}
Since the $Q$--values of the decays ${^{140}}{\rm Pr}^{58+} \to
{^{140}}{\rm Ce}^{58+} + e^+ + \nu_e$ and ${^{140}}{\rm Pr}^{57+} \to
{^{140}}{\rm Ce}^{57+} + e^+ + \nu_e$ are equal, $Q^{\rm H}_{\beta^+}
= Q^{\rm He}_{\beta^+} = Q_{\beta^+} = (3396 \pm 6)\,{\rm keV}$, the
Fermi integral $f(Q_{\beta^+},Z - 1) = (2.21 \pm 0.03)\,{\rm MeV}^5$
is defined by the phase volume of the final states of the decays and
the Fermi function, describing the Coulomb repulsion between the
positron and the nucleus ${^{140}}{\rm Ce}^{58+}$ for $Z = 59$
\cite{HS66}.

The theoretical values of the weak decay constants are defined up to
the unknown nuclear matrix element ${\cal M}_{\rm GT}$ \cite{GT} of
the Gamow--Teller transition, which cancels in the ratios.  The
theoretical ratios of the weak decay constants are given by
\begin{eqnarray}\label{label5}
  \hspace{-0.3in}&&R^{(\rm H),th}_{EC/\beta^+} = 
  \frac{3 \pi^2 Q^2_{\rm H} |\langle \psi^{(Z)}_{1s}\rangle|^2}{f(Q_{\beta^+},Z -1 )} = 
    1.40(4),\nonumber\\
    \hspace{-0.3in}&& R^{(\rm He),th}_{EC/\beta^+} = 
    \frac{2\pi^2 Q^2_{\rm He} 
      |\langle \psi^{(Z- 1)}_{1s}\psi^{(Z)}_{(1s)^2}\rangle|^2}{f(Q_{\beta^+},Z -1 )} =
 0.94(3),
    \nonumber\\
   \hspace{-0.3in}&& R^{\rm (H/He),th}_{EC/EC} =  \frac{2 I + 1}{2 F + 1}\,
\frac{|\langle \psi^{(Z)}_{1s}\rangle|^2}{|\langle \psi^{(Z- 1)}_{1s}\psi^{(Z)}_{(1s)^2}\rangle|^2}\,
    \frac{Q^2_{\rm H}}{Q^2_{\rm He}} = 1.50(4),
\end{eqnarray}
calculated for $F = \frac{1}{2}$ and $I = 1$. The experimental data on
the ratios of the weak decay constants are \cite{GSI}
\begin{eqnarray}\label{label6}
R^{(\rm H),\exp}_{EC/\beta^+} &=& 1.36(9),
\nonumber\\ R^{(\rm
  He),\exp}_{EC/\beta^+} &=& 0.96(8), \nonumber\\
 R^{\rm
  (H/He),\exp}_{EC/EC} &=& 1.49(8).
\end{eqnarray}
The theoretical values agree well with the experimental data
\cite{GSI}.

\section{Conclusion}

We have calculated the $EC$ and $\beta^+$ decay constants of the
H--like and He-like ions ${^{140}}{\rm Pr}^{58+}$ and ${^{140}}{\rm
  Pr}^{57+}$, respectively. Following the standard theory of weak
decays of heavy nuclei \cite{APR5} we have expressed the decay
constants in terms of the nuclear matrix element ${\cal M}_{\rm GT}$
of the Gamow--Teller transition. We have shown that the complete set
of experimental data on the ratios of weak decay constants of the ions
${^{140}}{\rm Pr}^{58+}$ and ${^{140}}{\rm Pr}^{57+}$ can be explained
within the standard theory of weak interactions of heavy ions.

Our theoretical values of the ratios of the decay constants for
H--like ${^{140}}{\rm Pr}^{58+}$ and He--like ${^{140}}{\rm Pr}^{57+}$
ions agree with the experimental data within an accuracy better than
$3\,\%$. This high precision of the theoretical analysis of the weak
decays of the H--like ${^{140}}{\rm Pr}^{58+}$ and He--like
${^{140}}{\rm Pr}^{57+}$ ions is due to the small number of electrons,
the behaviour of which can be described by the solution of the Dirac
equation. The dependence of the ratio $R_{EC/\beta^+}$ on the electron
structure of the decaying system is confirmed for both the
experimental data and our theoretical analysis of the He--like ion
${^{140}}{\rm Pr}^{57+}$. Indeed, for the He--like ${^{140}}{\rm
  Pr}^{57+}$ ion the ratio $R^{(\rm He),\exp}_{EC/\beta^+} = 0.96(8)$
is smaller than for the H--like ${^{140}}{\rm Pr}^{58+}$ ion $R^{(\rm
  H),\exp}_{EC/\beta^+} = 1.36(9)$.  Of course, such a regularity
should be confirmed experimentally by the measurements of $EC$ and
$\beta^+$ decays of the Li--like ${^{140}}{\rm Pr}^{56+}$ ions.

According to the hyperfine structure of the H--like ion ${^{140}}{\rm
  Pr}^{58+}$ \cite{HFS1}, the bound electron can be in two states with
a total angular momentum $F = \frac{1}{2}$ and $F = \frac{3}{2}$ with
the energy splitting equal to
\begin{eqnarray}\label{label7}
\Delta E = E_{1s_{F = \frac{1}{2}}} - E_{1s_{F = \frac{3}{2}}} = 
-\,2\alpha(\alpha Z)^3\,\frac{\mu}{\mu_N}\,\frac{m^2_e}{m_p}\,
\Big\{\frac{(1 - \delta)(1 - \varepsilon)}{(1 + \gamma)(1 + 2\gamma)}
 + x_{\rm rad}\Big\},
\end{eqnarray}
where $\gamma = \sqrt{1 - (\alpha Z)^2} - 1$, $\mu = + \,2.5\,\mu_N$
is the magnetic moment of the nucleus ${^{140}}{\rm Pr}^{59+}$
\cite{GSI}, $\mu_N = e/2m_p$ is the nuclear magneton, $m_p =
938.27\,{\rm MeV/c^2}$ is the proton mass, $\delta$ is the nuclear
charge distribution correction, $\varepsilon$ is the nuclear
magnetisation distribution correction (the Bohr--Weisskopf correction
\cite{HFS2}), $x_{\rm rad}$ denotes the radiative correction,
calculated to lowest order in $\alpha$ and $\alpha Z$ \cite{HFS3}.
Numerical estimates, carried our for different ions by Shabaev {\it et
al.} \cite{HFS1}, show that for the calculation of $\Delta E$ with an
accuracy better than $1\,\%$ one can drop the contributions of the
corrections $\delta$, $\varepsilon$ and $x_{\rm rad}$ and get $\Delta
E = -\,1,12,{\rm eV}$. The lifetime $\tau_{F = \frac{3}{2}}$ of the
hyperfine state of the H--like ${^{140}}{\rm Pr}^{58+}_{F =
\frac{3}{2}}$ ion is defined by the radiative transition ${^{140}}{\rm
Pr}^{58+}_{F = \frac{3}{2}} \to {^{140}}{\rm Pr}^{58+}_{F =
\frac{1}{2}} + \gamma$ only.  It is equal to $\tau_{F = \frac{3}{2}} =
8.5\times 10^{-3}\,{\rm sec}$ \cite{GSI}. Since the lifetime $\tau_{F
= \frac{3}{2}} = 8.5\times 10^{-3}\,{\rm sec}$ is much shorter than
the cooling time of about $2 s$ \cite{GSI}, all H--like ${^{140}}{\rm
Pr}^{58+}_{F = \frac{3}{2}}$ ions decay into the hyperfine ground
states ${^{140}}{\rm Pr}^{58+}_{F = \frac{1}{2}}$.

In our calculation of the $EC$--decay of the H--like ${^{140}}{\rm
Pr}^{58+}$ ion from the hyperfine ground state ${^{140}}{\rm
Pr}^{58+}_{F = \frac{1}{2}}$ the hyperfine structure is taken into
account in the spinorial wave function of the bound electron (see
Appendices A and B).

The influence of the hyperfine structure on the probabilities of weak
decays of heavy ions at finite temperature has been investigated by
Folan and Tsifrinovich \cite{HFS}. As has been mentioned by Folan and
Tsifrinovich \cite{HFS}, the $EC$--decay of the H--like ion
${^{140}}{\rm Pr}^{58+}_{F = \frac{3}{2}}$ from the hyperfine state
with $F = \frac{3}{2}$ is forbidden by a conservation of angular
momentum. This agrees with our analysis (see Appendix B).

Our result for the ratio $R^{\rm (H/He),th}_{EC/EC} = \frac{3}{2}$
agrees quantitatively well with that obtained by Patyk {\it et al.}
\cite{Patyk}, who also accounted for the hyperfine structure of the
H--like ${^{140}}{\rm Pr}^{58+}$ ion, and the estimate carried out by
Litvinov {\it et al.}  \cite{GSI}. According to Patyk {\it et al.}
\cite{Patyk} and Litvinov {\it et al.} \cite{GSI}, the ratio $R^{\rm
  (H/He),th}_{EC/EC} = \frac{3}{2}$ is fully caused by the statistical
factors
\begin{eqnarray}\label{label8}
R^{\rm (H/He)}_{EC/EC} =
\frac{2 I + 1}{2 F + 1} = \frac{3}{2},
\end{eqnarray}
calculated for $I = 1$ and $F = \frac{1}{2}$. Unfortunately, this
assertion is not completely correct. The result $R^{\rm
(H/He),th}_{EC/EC} = \frac{3}{2}$ appears only at the neglect of the
electron screening of the electric charge of the nucleus $Z$ in the
He--like ${^{140}}{\rm Pr}^{57+}$ ion. Having neglected the electron
screening of the electric charge of the nucleus $Z$ one can represent
the wave function of the bound $(1s)^2$ state of the He--like ion in
the form of the product of the one--electron Dirac wave functions. In
this case $\langle \psi^{(Z)}_{1s}\rangle = \langle \psi^{(Z-
1)}_{1s}\psi^{(Z)}_{(1s)^2}\rangle$ and the ratio $R^{\rm
(H/He),th}_{EC/EC}$, given by Eq.(\ref {label5}), reduces to
Eq.(\ref{label8}). Unlike, our analysis of the weak decays of the
H--like and He--like ions Patyk {\it et al.}  \cite{Patyk} as well as
Litvinov {\it et al.}  \cite{GSI} did not analyse the contributions of
the Coulomb wave functions of the bound electrons averaged over the
nuclear density.  Nevertheless, such contributions are important for
the correct calculation of both the ratio $R^{\rm (H/He)}_{EC/EC}$ for
ions with small electric charges $Z$ and the ratios $R^{(\rm
H)}_{EC/\beta^+}$ and $R^{(\rm He)}_{EC/\beta^+}$ (see
Eq.(\ref{label5})), which were calculated by neither Patyk {\it et
al.} \cite{Patyk} nor Litvinov {\it et al.}  \cite{GSI}.

\section*{Acknowledgement}

We acknowledge many fruitful discussions with M. Kleber, F. Bosch and
Yu. A. Litvinov in the course of this work.  One of the authors (A.
Ivanov) is greatful to N. I. Troitskaya and V. M. Shabaev for the
discussions of properties of heavy ions.

\newpage

\section*{Appendix A: Calculation of weak decay constants of H-like
 ${^{140}}{\rm Pr}^{58+}$ and He--like ${^{140}}{\rm Pr}^{57+}$ ions}
\renewcommand{\theequation}{A-\arabic{equation}}
\setcounter{equation}{0}

In this Appendix we adduce the detailed calculations of the $EC$ and
$\beta^+$ decay constants of the H--like ${^{140}}{\rm Pr}^{58+}$ ion
in the hyperfine ground state ${^{140}}{\rm Pr}^{58+}_{F =
\frac{1}{2}}$ and the He--like ${^{140}}{\rm Pr}^{57+}_{I = 1}$ ion in
the ground $(1s)^2$ state. For the calculation of the weak decay
constants of H-like and He--like ions we follow \cite{APR5} and use
the Hamiltonian Eq.(\ref{label1}).

\subsubsection*{$EC$--decay of the H--like ${^{140}}{\rm Pr}^{58+}$ ion}

The K--shell electron capture decay (the $EC$--decay) ${^{140}}{\rm
Pr}^{58+}_{F = \frac{1}{2}}\to {^{140}}{\rm Ce}^{58+}_{I = 0} + \nu_e$
describes a transition of the H--like mother ion ${^{140}}{\rm
Pr}^{58+}$ from the hyperfine ground state $|F,M_F\rangle$ with $F =
1/2$ and $M_F = \pm 1/2$ into the daughter ion ${^{140}}{\rm
Ce}^{58+}$ in the ground state $|I,I_z\rangle = |0,0\rangle$. The
$EC$--decay constant of the H--like ${^{140}}{\rm Pr}^{58+}$ ion is
defined by
\begin{eqnarray}\label{labelA.1} 
  \lambda^{(\rm H)}_{EC} = \frac{1}{2M_m}\,\frac{1}{2 F + 1}\sum_{M_F}
\int |M^{EC}_{_{F,M_F}}|^2\,(2\pi)^4\delta^{(4)}(p_d + k - p_m)\,
\frac{d^3p_d}{(2\pi)^3 2E_d}\frac{d^3k}{(2\pi)^3 2E_{\nu}},
\end{eqnarray}
where $p_m$, $p_d$ and $k$ are 4--momenta of the mother ion, the
daughter ion and the neutrino, respectively.  According to
\cite{APR5}, the amplitudes $M^{EC}_{F,M_F}$ of the $EC$--decay are
equal to
\begin{eqnarray}\label{labelA.2} 
  \hspace{-0.3in} M^{EC}_{\frac{1}{2},+\frac{1}{2}} &=& \sqrt{2 M_m 2E_d E_{\nu}}\,
  {\cal M}^{EC}_{\rm GT}\,\Bigg\{\sqrt{\frac{2}{3}}\,[\varphi^{\dagger}_{n,-\frac{1}{2}}
  \vec{\sigma}\,\varphi_{p,+\frac{1}{2}}]\cdot 
[\varphi^{\dagger}_{\nu,-\frac{1}{2}}(1 - \vec{n}\cdot\vec{\sigma}\,)
  \vec{\sigma}\,\varphi_{e,-\frac{1}{2}}]\nonumber\\
 \hspace{-0.3in}&& - \sqrt{\frac{1}{6}}\,[\varphi^{\dagger}_{n,+\frac{1}{2}}
  \vec{\sigma}\,\varphi_{p,+\frac{1}{2}} - \varphi^{\dagger}_{n,-\frac{1}{2}}
  \vec{\sigma}\,\varphi_{p,-\frac{1}{2}}]\cdot 
[\varphi^{\dagger}_{\nu,-\frac{1}{2}}(1 - \vec{n}\cdot\vec{\sigma}\,)
  \vec{\sigma}\,\varphi_{e,+\frac{1}{2}}]\Bigg\},\nonumber\\
  \hspace{-0.3in} M^{EC}_{\frac{1}{2},-\frac{1}{2}} &=& \sqrt{2 M_m 2E_d E_{\nu}}\,{\cal M}^{EC}_{\rm GT}
\,\Bigg\{\sqrt{\frac{2}{3}}\,[\varphi^{\dagger}_{n,+\frac{1}{2}}
  \vec{\sigma}\,\varphi_{p,-\frac{1}{2}}]\cdot 
[\varphi^{\dagger}_{\nu,-\frac{1}{2}}(1 - \vec{n}\cdot\vec{\sigma}\,)
  \vec{\sigma}\,\varphi_{e,+\frac{1}{2}}]\nonumber\\
 \hspace{-0.3in}&& - \sqrt{\frac{1}{6}}\,[\varphi^{\dagger}_{n,+\frac{1}{2}}
  \vec{\sigma}\,\varphi_{p,+\frac{1}{2}} - \varphi^{\dagger}_{n,-\frac{1}{2}}
  \vec{\sigma}\,\varphi_{p,-\frac{1}{2}}]\cdot 
[\varphi^{\dagger}_{\nu,-\frac{1}{2}}(1 - \vec{n}\cdot\vec{\sigma}\,)
  \vec{\sigma}\,\varphi_{e,-\frac{1}{2}}]\Bigg\},
\end{eqnarray}
where we have used the non--relativistic approximation for nucleons
\cite{APR5}, $\sqrt{2/3}$ and $\sqrt{1/6}$ are the Clebsch--Gordon
coefficients of the spinorial wave function of the bound electron in
the hyperfine state with $F = \frac{1}{2}$, caused by the spinorial
wave functions of the nucleus ${^{140}}{\rm Pr}^{59+}$ with spin $I =
1$ and the electron with spin $s = \frac{1}{2}$, $\vec{n} =
\vec{k}/E_{\nu}$ is a unit vector alone the 3--momentum of the
neutrino and ${\cal M}^{EC}_{\rm GT}$ is the matrix element of the
Gamow--Teller transition defined by
\begin{eqnarray}\label{labelA.3} 
{\cal M}^{EC}_{\rm GT} = -\,g_A\,\frac{G_F}{\sqrt{2}}\,V_{ud}\int d^3x\,
\Psi^{\,*}_d(\vec{r}\,)\Psi_m(\vec{r}\,)\,\psi^{(Z)}_{1s}(r),
\end{eqnarray}
where $\Psi^{\,*}_d(\vec{r}\,)$ and $\Psi_m(\vec{r}\,)$ are the wave
functions of the daughter ${^{140}}{\rm Ce}^{58+}$ and mother
${^{140}}{\rm Pr}^{59+}$ nuclei, respectively, and
$\psi^{(Z)}_{1s}(r)$ is the radial wave function of the bound electron
in the hyperfine ground state of the H--like ${^{140}}{\rm
Pr}^{58+}_{F = \frac{1}{2}}$ ion with electric charge $Z$. It is equal
to \cite{IZ80}
\begin{eqnarray}\label{labelA.4} 
 \hspace{-0.3in} \psi^{(Z)}(r) =
  \sqrt{\frac{1}{\pi a^3_B}\,\frac{2 + \gamma}{\Gamma(3 + 2\gamma)}}\,
    \Big(\frac{2r}{a_B}\Big)^{\gamma}\,e^{\textstyle \,-\,r/a_B},
\end{eqnarray}
where $\gamma = \sqrt{1 - \alpha^2 Z^2 } - 1$ \cite{IZ80}. For
numerical calculations we set
$\Psi^{\,*}_d(\vec{r}\,)\Psi_m(\vec{r}\,) \sim \rho(r)$, where
$\rho(r)$ has the Woods--Saxon shape \cite{GT}. For the
subsequent analysis it is convenient to rewrite the matrix element
Eq.(\ref{labelA.3}) as follows
\begin{eqnarray}\label{labelA.5} 
{\cal M}^{EC}_{\rm GT} = \frac{1}{2\sqrt{2}}\,{\cal M}_{\rm GT}\,
\langle \psi^{(Z)}_{1s}\rangle, 
\end{eqnarray}
where we have denoted
\begin{eqnarray}\label{labelA.6} 
{\cal M}_{\rm GT} = -\,2\,g_A\,G_F\,V_{ud}\int d^3x\,
\rho(r)\;,\; \langle \psi^{(Z)}_{1s}\rangle = 
\frac{\displaystyle \int d^3x\,\rho(r)
\psi^{(Z)}_{1s}(\vec{r}\,)}{\displaystyle \int d^3x\,
\rho(r)}.
\end{eqnarray}
Since a neutrino is polarised anti--parallel to its spin, i.e. the
spinorial wave function is the eigenfunction of the operator
$\vec{n}\cdot\vec{\sigma}$ with the eigenvalue $-\,1$ (see \cite{HS66,EK66})
\begin{eqnarray}\label{labelA.7} 
\vec{n}\cdot\vec{\sigma}\,\varphi_{\nu,-\frac{1}{2}} = -\,\varphi_{\nu,-\frac{1}{2}},
\end{eqnarray}
a non--zero contribution to the $EC$--decay constant comes from the
amplitude $M^{EC}_{\frac{1}{2},-\frac{1}{2}}$ only
\begin{eqnarray}\label{labelA.8} 
  M^{EC}_{\frac{1}{2},-\frac{1}{2}} &=& \frac{1}{\sqrt{2}}\,\sqrt{2 M_m 2E_d E_{\nu}}
\,{\cal M}_{\rm GT}\,\langle \psi^{(Z)}_{1s}\rangle\,
\Bigg\{\sqrt{\frac{2}{3}}\,[\varphi^{\dagger}_{n,+\frac{1}{2}}
\vec{\sigma}\,\varphi_{p,-\frac{1}{2}}]\cdot [\varphi^{\dagger}_{\nu,-\frac{1}{2}}
\vec{\sigma}\,\varphi_{e,+\frac{1}{2}}],\nonumber\\
 \hspace{-0.3in}&-& \sqrt{\frac{1}{6}}\,[\varphi^{\dagger}_{n,+\frac{1}{2}}
  \vec{\sigma}\,\varphi_{p,+\frac{1}{2}} - \varphi^{\dagger}_{n,-\frac{1}{2}}
  \vec{\sigma}\,\varphi_{p,-\frac{1}{2}}]\cdot [\varphi^{\dagger}_{\nu,-\frac{1}{2}}
  \vec{\sigma}\,\varphi_{e,-\frac{1}{2}}]\Bigg\},
\end{eqnarray}
where we have taken into account Eq.(\ref{labelA.7}). Since the
spinorial matrix elements are equal to
\begin{eqnarray}\label{labelA.9} 
 { [\varphi^{\dagger}_{n,+\frac{1}{2}}
  \vec{\sigma}\,\varphi_{p,-\frac{1}{2}}]\cdot [\varphi^{\dagger}_{\nu,-\frac{1}{2}}
  \vec{\sigma}\,\varphi_{e,+\frac{1}{2}}]} &=& +\,2,\nonumber\\
  {[\varphi^{\dagger}_{n,+\frac{1}{2}}
  \vec{\sigma}\,\varphi_{p,+\frac{1}{2}} - \varphi^{\dagger}_{n,-\frac{1}{2}}
  \vec{\sigma}\,\varphi_{p,-\frac{1}{2}}]\cdot [\varphi^{\dagger}_{\nu,-\frac{1}{2}}
  \vec{\sigma}\,\varphi_{e,-\frac{1}{2}}]} &=&- \,2,
\end{eqnarray}
the amplitude $M^{EC}_{\frac{1}{2},-\frac{1}{2}}$ is
\begin{eqnarray}\label{labelA.10} 
 \hspace{-0.3in} M^{EC}_{\frac{1}{2},-\frac{1}{2}} = \sqrt{3}\,\sqrt{2 M_m 2E_d E_{\nu}}
\,{\cal M}^{EC}_{\rm GT}\,\langle \psi^{(Z)}_{1s}\rangle.
\end{eqnarray}
Taking the squared absolute value of the amplitude
Eq.(\ref{labelA.10}) and substituting into Eq.(\ref{labelA.1}) we get
\begin{eqnarray}\label{labelA.11} 
  \lambda^{(\rm H)}_{EC} &=& \frac{3}{2 F + 1}\, |{\cal M}_{\rm GT}|^2
  |\langle \psi^{(Z)}_{1s}\rangle|^2 \int \delta^{(4)}(p_d + k -
  p_m)\, \frac{d^3p_d d^3k}{8\pi^2} = \nonumber\\ &=& \frac{1}{2 F +
  1}\,\frac{3}{2}\,|{\cal M}_{\rm GT}|^2\,|\langle
  \psi^{(Z)}_{1s}\rangle|^2\, \frac{Q^2_{\rm H}}{\pi},
\end{eqnarray}
where in the ratio $\frac{3}{2}$ the factors 3 and $\frac{1}{2}$ are
caused by the hyperfine structure of the ground state of the H--like
${^{140}}{\rm Pr}^{58+}_{F = \frac{1}{2}}$ ion and the phase volume of the
final state of the decay, respectively.

Thus, the $EC$--decay constant of the H--like ${^{140}}{\rm
Pr}^{58+}$ ion is
\begin{eqnarray}\label{labelA.12} 
  \lambda^{(\rm H)}_{EC} = \frac{1}{2 F + 1}\,\frac{3}{2}\,|{\cal M}_{\rm GT}|^2\,
|\langle \psi^{(Z)}_{1s}\rangle|^2\,
\frac{Q^2_{\rm H}}{\pi},
\end{eqnarray}
where $Q_{\rm H} = (3348\pm 6)\,{\rm keV}$ is the $Q$--value of the
$EC$--decay of the H--like ${^{140}}{\rm Pr}^{58+}$ ion.

\subsubsection*{$\beta^+$--decay of H--like ${^{140}}{\rm Pr}^{58+}$ ion}

The $\beta^+$--decay ${^{140}}{\rm Pr}^{58+}_{F = \frac{1}{2}}\to
{^{140}}{\rm Ce}^{57+}_{F' = \frac{1}{2}} + e^+ + \nu_e$ describes a
transition of the H--like mother ion ${^{140}}{\rm Pr}^{58+}_{F =
\frac{1}{2}}$ from the hyperfine ground state $|F,M_F\rangle$ with $F
= 1/2$ and $M_F = \pm 1/2$ into the H--like daughter ion ${^{140}}{\rm
Ce}^{57+}_{F' = \frac{1}{2}}$ in the ground state $|F',M'_F\rangle$
with $F' = 1/2$ and $M'_F = \pm 1/2$. The $\beta^+$--decay constant of
the H--like ${^{140}}{\rm Pr}^{58+}$ ion is defined by
\begin{eqnarray}\label{labelA.13} 
  \lambda^{(\rm H)}_{\beta^+} &=& \frac{1}{2M_m}\,\frac{1}{2 F + 1}\sum_{M_F,M'_F}\int 
  |M^{\beta^+}_{_{F,M_F \to F',M'_F }}|^2\,
  F(Z-1,E_+)\nonumber\\
  &&\times\,(2\pi)^4\delta^{(4)}(p_d + k + p_+ - p_m)\,
  \frac{d^3p_d}{(2\pi)^3 2E_d}\frac{d^3k}{(2\pi)^3 2E_{\nu}}
  \frac{d^3p_+}{(2\pi)^3 2E_+},
\end{eqnarray}
where $p_m$, $p_d$, $k$ and $p_+$ are 4--momenta of the mother ion,
the daughter ion, the neutrino and the positron, respectively, $F(Z -
1,E_+)$ is the Fermi function \cite{HS66}. 

The amplitudes of the $\beta^+$--decay $M^{\beta^+}_{_{F,M_F \to
F',M'_F} }$ are equal to
\begin{eqnarray}\label{labelA.14} 
  \hspace{-0.3in}&&M^{\beta^+}_{\frac{1}{2},+\frac{1}{2}\to \frac{1}{2},M'_F } = \sqrt{2 M_m 2E_d }\,
  {\cal M}^{\beta^+}_{\rm GT}\,\Bigg\{\sqrt{\frac{2}{3}}\,[\varphi^{\dagger}_{n,-\frac{1}{2}}
  \vec{\sigma}\,\varphi_{p,+\frac{1}{2}}]\cdot [\bar{u}_{\nu}(\vec{k},-\frac{1}{2})\vec{\gamma}\,
  (1 - \gamma^5)\,v_{e^+}(\vec{p}_+,\sigma_+)]\nonumber\\
  \hspace{-0.3in}&&\times \delta_{M'_F,-\frac{1}{2}} - \sqrt{\frac{1}{6}}\,[\varphi^{\dagger}_{n,+\frac{1}{2}}
  \vec{\sigma}\,\varphi_{p,+\frac{1}{2}} - \varphi^{\dagger}_{n,-\frac{1}{2}}
  \vec{\sigma}\,\varphi_{p,-\frac{1}{2}}]\cdot [\bar{u}_{\nu}(\vec{k},-\frac{1}{2})\vec{\gamma}\,
  (1 - \gamma^5)\,v_{e^+}(\vec{p}_+,\sigma_+)] \delta_{M'_F,+\frac{1}{2}}\Bigg\},\nonumber\\
 \hspace{-0.3in} && M^{\beta^+}_{\frac{1}{2},-\frac{1}{2}\to \frac{1}{2},M'_F} = 
\sqrt{2 M_m 2E_d}\, {\cal M}^{\beta^+}_{\rm GT}
  \Bigg\{\sqrt{\frac{2}{3}}\,
  [\bar{u}_{\nu}(\vec{k},-\frac{1}{2})(\gamma_1 + i\gamma_2)\,(1 - \gamma^5)
\,v_{e^+}(\vec{p}_+,\sigma_+)]\delta_{M'_F,+\frac{1}{2}}
  \nonumber\\
  \hspace{-0.3in}&&\times \delta_{M'_F,+\frac{1}{2}}  - 
\sqrt{\frac{1}{6}}\,[\varphi^{\dagger}_{n,+\frac{1}{2}}
  \vec{\sigma}\,\varphi_{p,+\frac{1}{2}} - \varphi^{\dagger}_{n,-\frac{1}{2}}
  \vec{\sigma}\,\varphi_{p,-\frac{1}{2}}]\cdot [\bar{u}_{\nu}(\vec{k},-\frac{1}{2})\vec{\gamma}\,
  (1 - \gamma^5)\,v_{e^+}(\vec{p}_+,\sigma_+)] \delta_{M'_F,-\frac{1}{2}} \Bigg\},\nonumber\\
  \hspace{-0.3in}&&
\end{eqnarray}
where $M'_F = \pm 1/2$ defines a polarisation of the electron in the
final state. The matrix element of the Gamow--Teller transition ${\cal
M}^{\beta^+}_{\rm GT}$ is defined by
\begin{eqnarray}\label{labelA.15} 
  {\cal M}^{\beta^+}_{\rm GT} = -\,g_A\,\frac{G_F}{\sqrt{2}}\,V_{ud}
\langle \psi^{(Z-1)}_{1s}|\psi^{(Z)}_{1s}\rangle
\int d^3x\,\rho(r) = \frac{1}{2\sqrt{2}}\,{\cal M}_{\rm GT}\,
\langle \psi^{(Z-1)}_{1s}|\psi^{(Z)}_{1s}\rangle,
\end{eqnarray}
where $\langle \psi^{(Z-1)}_{1s}|\psi^{(Z)}_{1s}\rangle$ is the matrix
element of the transition of the bound electron from the initial state
with the wave function $\psi^{(Z)}_{1s}$ into the final state with the
wave function $\psi^{(Z-1)}_{1s}$.  Since for $Z = 59$ this matrix
element is equal to unity with a good accuracy, we set $\langle
\psi^{(Z-1)}_{1s}|\psi^{(Z)}_{1s}\rangle = 1$ and $\tilde{\cal
M}^{\beta^+}_{\rm GT} = {\cal M}_{\rm GT}/2\sqrt{2}$, where ${\cal
M}_{\rm GT}$ is defined by Eq.(\ref{labelA.6}). Having calculated the
proton--neutron matrix elements
\begin{eqnarray}\label{labelA.16}
\hspace{-0.3in}  {[\varphi^{\dagger}_{n,-\frac{1}{2}}
  \sigma_j\,\varphi_{p,+\frac{1}{2}}]}&=& \delta_{1j} - i\,\delta_{2j},\nonumber\\
  \hspace{-0.3in}  {[\varphi^{\dagger}_{n,+\frac{1}{2}}
  \sigma_j\,\varphi_{p,-\frac{1}{2}}]}&=& \delta_{1j} + i\,\delta_{2j},\nonumber\\
 \hspace{-0.3in}{[\varphi^{\dagger}_{n,+\frac{1}{2}}
  \sigma_j\,\varphi_{p,+\frac{1}{2}} - \varphi^{\dagger}_{n,-\frac{1}{2}}
  \sigma_j\,\varphi_{p,-\frac{1}{2}}]} &=&2\,\delta_{3j}
\end{eqnarray}
we get
\begin{eqnarray}\label{labelA.17} 
 \hspace{-0.3in}&&M^{\beta^+}_{\frac{1}{2},+\frac{1}{2}\to \frac{1}{2},M'_F } = \sqrt{2 M_m 2E_d }\,
  \frac{{\cal M}_{\rm GT}}{2\sqrt{2}}\,\sqrt{\frac{2}{3}}\Big\{[\bar{u}_{\nu}(\vec{k},-\frac{1}{2})
(\gamma_1 - i \gamma_2)\,
  (1 - \gamma^5)\,v_{e^+}(\vec{p}_+,\sigma_+)]\,\delta_{M'_F,-\frac{1}{2}}\nonumber\\
  \hspace{-0.3in}&&  - [\bar{u}_{\nu}(\vec{k},-\frac{1}{2})\gamma_3\,
  (1 - \gamma^5)\,v_{e^+}(\vec{p}_+,\sigma_+)]\, \delta_{M'_F,+\frac{1}{2}}\Big\},\nonumber\\
 \hspace{-0.3in} && M^{\beta^+}_{\frac{1}{2},-\frac{1}{2}\to \frac{1}{2},M'_F} = \sqrt{2 M_m 2E_d }\,
  \frac{{\cal M}_{\rm GT}}{2\sqrt{2}}\,\sqrt{\frac{2}{3}}
  \Big\{
  [\bar{u}_{\nu}(\vec{k},-\frac{1}{2})(\gamma_1 + i\gamma_2)\,
(1 - \gamma^5)\,v_{e^+}(\vec{p}_+,\sigma_+)]\delta_{M'_F,+\frac{1}{2}} 
  \nonumber\\
  \hspace{-0.3in}&&- [\bar{u}_{\nu}(\vec{k},-\frac{1}{2})\gamma_3\,
  (1 - \gamma^5)\,v_{e^+}(\vec{p}_+,\sigma_+)] \,\delta_{M'_F,-\frac{1}{2}} \Big\},
\end{eqnarray}
The squared absolute values of the amplitudes, summed over the
polarisations of the final state are equal to
\begin{eqnarray}\label{labelA.18} 
 \hspace{-0.3in} \sum_{M'_F =\pm 1/2}|M^{\beta^+}_{\frac{1}{2},+\frac{1}{2}\to \frac{1}{2},M'_F }|^2 &=& 2 M_m 2E_d\,
\frac{|{\cal M}_{\rm GT}|^2}{8}
  \,\frac{2}{3}\,\Big\{{\rm tr}\{\hat{k}\gamma_3(1 - \gamma^5)
  (\hat{p}_+ - m_e)\gamma_3(1 - \gamma^5)\}\nonumber\\
&+& {\rm tr}\{\hat{k}(\gamma_1 + i\gamma_2)(1 - \gamma^5)
  (\hat{p}_+ - m_e)(\gamma_1 - i\gamma_2)(1 - \gamma^5)\}\Big\}\nonumber\\
  \hspace{-0.3in} \sum_{M'_F =\pm 1/2} |M^{\beta^+}_{\frac{1}{2},-\frac{1}{2}\to \frac{1}{2},M'_F}|^2 &=& 
2 M_m 2E_d\,\frac{|{\cal M}_{\rm GT}|^2}{8}
  \,\frac{2}{3}\,\Big\{{\rm tr}\{\hat{k}\gamma_3(1 - \gamma^5)
  (\hat{p}_+ - m_e)\gamma_3(1 - \gamma^5)\}\nonumber\\
 \hspace{-0.3in}&+& {\rm tr}\{\hat{k}(\gamma_1 - i\gamma_2)(1 - \gamma^5)
  (\hat{p}_+ - m_e)(\gamma_1 + i\gamma_2)(1 - \gamma^5)\}\Big\}.
\end{eqnarray}
The results of the calculation of the traces over Dirac matrices are
given by
\begin{eqnarray}\label{labelA.19} 
  \hspace{-0.3in} &&{\rm tr}\{\hat{k}(\gamma_1 - i\gamma_2)(1 - \gamma^5)
  (\hat{p}_+ - m_e)(\gamma_1 + i\gamma_2)(1 - \gamma^5)\} = 
  2\,{\rm tr}\{\hat{k}(\gamma_1 -i\gamma_2)
  \hat{p}_+ (\gamma_1 + i\gamma_2)\} = \nonumber\\
  \hspace{-0.3in}&&= 2{\rm tr}\{\hat{k}\gamma_1\hat{p}_+\gamma_1\} 
  + 2{\rm tr}\{\hat{k}\gamma_2\hat{p}_+\gamma_2\}  + 2i{\rm tr}\{\hat{k}\gamma_1\hat{p}_+\gamma_2\} 
  - 2i{\rm tr}\{\hat{k}\gamma_2\hat{p}_+\gamma_1\} =\nonumber\\
  \hspace{-0.3in}&&= 8\,(2 k_x p_{+x} + k\cdot p_+) + 8\,(2 k_y p_{+y} + k\cdot p_+) 
  + 8i\,( k_x p_{+y} + k_y p_{+x}) - 
  8i\,( k_y p_{+x} + k_x p_{+y})  = \nonumber\\
  \hspace{-0.3in}&&= 16\,(E_{\nu} E_+ - k_z p_{+z}),\nonumber\\
  \hspace{-0.3in} &&{\rm tr}\{\hat{k}(\gamma_1 + i\gamma_2)(1 - \gamma^5)
  (\hat{p}_+ - m_e)(\gamma_1 -  i\gamma_2)(1 - \gamma^5)\} =\nonumber\\
  \hspace{-0.3in}&&=  
  {\rm tr}\{\hat{k}(\gamma_1 - i\gamma_2)(1 - \gamma^5)
  (\hat{p}_+ - m_e)(\gamma_1 + i\gamma_2)(1 - \gamma^5)\}  =  16\,(E_{\nu} E_+ - k_z p_{+z}),\nonumber\\
  \hspace{-0.3in} &&{\rm tr}\{\hat{k}\gamma_3(1 - \gamma^5)
  (\hat{p}_+ - m_e)\gamma_3(1 - \gamma^5)\} = 8\,(E_{\nu}E_+ + k_z p_{+z}).
\end{eqnarray}
Using Eq.(\ref{labelA.19}), for the sum of the squared absolute values
of the amplitudes Eq.(\ref{labelA.18}) we obtain the following
expression
\begin{eqnarray}\label{labelA.20} 
 \hspace{-0.3in} \sum_{M_F, M'_F =\pm 1/2}|M^{\beta^+}_{F, M_F\to F',M'_F }|^2 &=& 2 M_m 2E_d\,
|{\cal M}_{\rm GT}|^2\,4\,(E_{\nu}E_+ - \frac{1}{3}\,k_z p_{+z}).
\end{eqnarray}
Substituting Eq.(\ref{labelA.20}) into Eq.(\ref{labelA.13}) and
integrating over the phase volume of the final state we get
\begin{eqnarray}\label{labelA.21}
  \hspace{-0.3in}  
\lambda^{(\rm H)}_{\beta^+} &=& \frac{1}{2 F + 1}\,|{\cal M}_{\rm GT}|^2\int
  F(Z - 1, E_+)\,\delta^{(4)}(p_d + k + p_+ - p_m)\,
  \frac{d^3p_dd^3kd^3p_+}{32\pi^5} =\nonumber\\
 \hspace{-0.3in} &=& \frac{2}{2 F + 1}\,
\frac{|{\cal M}_{\rm GT}|^2}{4\pi^3}\,
f(Q_{\beta^+}, Z - 1), 
\end{eqnarray}
where $f(Q_{\beta^+}, Z - 1)$ is defined by \cite{HS66}
\begin{eqnarray}\label{labelA.22} 
 f(Q_{\beta^+}, Z - 1) = \int^{Q_{\beta^+} - m_e}_{m_e}(Q_{\beta^+} - m_e - E_+)^2\sqrt{E^2_+ - m^2_e}\,
F(Z - 1,E_+)\, E_+ dE_+ .
\end{eqnarray}
Thus, the $\beta^+$--decay constant of the H--like ${^{140}}{\rm
Pr}^{58+}$ ion is equal to
\begin{eqnarray}\label{labelA.23} 
  \lambda^{(\rm H)}_{\beta^+} = \frac{2}{2 F + 1}\,\frac{|{\cal M}_{\rm GT}|^2}{4\pi^3}\,
f(Q_{\beta^+}, Z - 1)
\end{eqnarray}
and $Q_{\beta^+} = (3396\pm 6)\,{\rm keV}$ is the $Q$--value of the
$\beta^+$--decay of the H--like ${^{140}}{\rm Pr}^{58+}$ ion.

\subsubsection*{$EC$--decay of He--like ${^{140}}{\rm Pr}^{57+}$ ion}

The K--shell electron capture decay (the $EC$--decay) ${^{140}}{\rm
Pr}^{57+}_{I = 1}\to {^{140}}{\rm Ce}^{57+}_{F = \frac{1}{2}} + \nu_e$
describes a transition of the He--like mother ion ${^{140}}{\rm
Pr}^{57+}$ from the ground $(1s)^2$ state $|I,I_z\rangle$ with $I = 1$
and $I_z = 0,\pm 1$ into the H--like daughter ion ${^{140}}{\rm
Ce}^{57+}$ in the ground state $|F, M_F\rangle$ with $F = 1/2$ and
$M_F = \pm 1/2$. The $EC$--decay constant is defined by
\begin{eqnarray}\label{labelA.24} 
  \lambda^{(\rm He)}_{EC} &=& 
\frac{1}{2M_m}\,\frac{1}{2 I + 1}\sum_{I_z,M_F}\int |M^{EC}_{_{I,I_z \to F,M_F}}|^2\,
(2\pi)^4\delta^{(4)}(p_d + k - p_m)\,
\frac{d^3p_d}{(2\pi)^3 2E_d}\frac{d^3k}{(2\pi)^3 2E_{\nu}}.\nonumber\\
&&
\end{eqnarray}
The amplitudes of the $EC$--decay $M^{EC}_{I,I_z \to F,M_F}$ are
equal to
\begin{eqnarray}\label{labelA.25} 
 \hspace{-0.3in} M^{EC}_{1,+1\to \frac{1}{2},+\frac{1}{2}} &=& \sqrt{2 M_m 2E_d E_{\nu}}\,
\tilde{\cal M}^{EC}_{\rm GT}\,[\varphi^{\dagger}_{n,-\frac{1}{2}}
\vec{\sigma}\,\varphi_{p,+\frac{1}{2}}]\cdot [\varphi^{\dagger}_{\nu,-\frac{1}{2}}(1 - \vec{n}\cdot\vec{\sigma}\,)
\vec{\sigma}\,\varphi_{e,-\frac{1}{2}}],\nonumber\\
\hspace{-0.3in} M^{EC}_{1,+1\to \frac{1}{2},-\frac{1}{2}} &=& \sqrt{2 M_m 2E_d E_{\nu}}\,
\tilde{\cal M}^{EC}_{\rm GT}\,[\varphi^{\dagger}_{n,-\frac{1}{2}}
\vec{\sigma}\,\varphi_{p,+\frac{1}{2}}]\cdot [\varphi^{\dagger}_{\nu,-\frac{1}{2}}(1 - \vec{n}\cdot\vec{\sigma}\,)
\vec{\sigma}\,\varphi_{e,+\frac{1}{2}}],\nonumber\\
 \hspace{-0.3in} M^{EC}_{1,0\to \frac{1}{2},+\frac{1}{2}} &=& \sqrt{2 M_m 2E_d  E_{\nu}}\,
\tilde{\cal M}^{EC}_{\rm GT}\,\frac{1}{\sqrt{2}}\,[\varphi^{\dagger}_{n,+\frac{1}{2}}
\vec{\sigma}\,\varphi_{p,+\frac{1}{2}} - \varphi^{\dagger}_{n,-\frac{1}{2}}
\vec{\sigma}\,\varphi_{p,-\frac{1}{2}}]\nonumber\\
\hspace{-0.3in}&&\cdot [\varphi^{\dagger}_{\nu,-\frac{1}{2}}(1 - \vec{n}\cdot\vec{\sigma}\,)
\vec{\sigma}\,\varphi_{e,-\frac{1}{2}}],\nonumber\\
 \hspace{-0.3in} M^{EC}_{1,0\to \frac{1}{2},-\frac{1}{2}} &=& \sqrt{2 M_m 2E_d  E_{\nu}}\,\tilde{\cal M}^{EC}_{\rm GT}
\,\frac{1}{\sqrt{2}}\,[\varphi^{\dagger}_{n,+\frac{1}{2}}
\vec{\sigma}\,\varphi_{p,+\frac{1}{2}} - \varphi^{\dagger}_{n,-\frac{1}{2}}
\vec{\sigma}\,\varphi_{p,-\frac{1}{2}}]\nonumber\\
\hspace{-0.3in}&&\cdot [\varphi^{\dagger}_{\nu,-\frac{1}{2}}(1 - \vec{n}\cdot\vec{\sigma}\,)
\vec{\sigma}\,\varphi_{e,+\frac{1}{2}}],\nonumber\\
 \hspace{-0.3in} M^{EC}_{1,-1 \to \frac{1}{2},+\frac{1}{2}} &=& \sqrt{2 M_m 2E_d E_{\nu}}\,
\tilde{\cal M}^{EC}_{\rm GT}\,
[\varphi^{\dagger}_{n,+\frac{1}{2}}
\vec{\sigma}\,\varphi_{p,-\frac{1}{2}}]\cdot [\varphi^{\dagger}_{\nu,-\frac{1}{2}}(1 - \vec{n}\cdot\vec{\sigma}\,)
\vec{\sigma}\,\varphi_{e,-\frac{1}{2}}],\nonumber\\
\hspace{-0.3in} M^{EC}_{1,-1 \to \frac{1}{2},-\frac{1}{2}} &=& \sqrt{2 M_m 2E_d E_{\nu}}\,
\tilde{\cal M}^{EC}_{\rm GT}\,
[\varphi^{\dagger}_{n,+\frac{1}{2}}
\vec{\sigma}\,\varphi_{p,-\frac{1}{2}}]\cdot [\varphi^{\dagger}_{\nu,-\frac{1}{2}}(1 - \vec{n}\cdot\vec{\sigma}\,)
\vec{\sigma}\,\varphi_{e,+\frac{1}{2}}],\nonumber\\
 \hspace{-0.3in}&&
\end{eqnarray}
where $\vec{n} = \vec{k}/E_{\nu}$ is a unit vector alone the
3--momentum of the neutrino and we have denoted
\begin{eqnarray}\label{labelA.26} 
  \tilde{\cal M}^{EC}_{\rm GT} &=& \frac{{\cal M}_{\rm GT}}{2\sqrt{2}}\,
\langle \psi^{(Z-1)}_{1s} \psi^{(Z)}_{(1s)^2}\rangle,\nonumber\\
  \langle \psi^{(Z-1)}_{1s} \psi^{(Z)}_{(1s)^2}\rangle &=& 
  \frac{\int\!\! d^3x_1d^3x_2\psi^{(Z-1)}_{1s}(\vec{r}_1)\psi^{\rm
      (Z)}_{(1s)^2}(\vec{r}_1,\vec{r}_2)\rho(r_2)}{\int\! d^3x_2\rho(r_2)},
\end{eqnarray}
where $\psi^{\rm (Z)}_{(1s)^2}(\vec{r}_1,\vec{r}_2)$ is the wave
      function of the ground $(1s)^2$ state of the He--like
      ${^{140}}{\rm Pr}^{57+}$ ion and ${\cal M}_{\rm GT}$ is the
      matrix element of the Gamow--Teller transition defined by
      Eq.(\ref{labelA.6}).

For the amplitudes of the $EC$--decay of the He--like ${^{140}}{\rm
      Pr}^{57+}$ ion we get
\begin{eqnarray}\label{labelA.27} 
 \hspace{-0.3in} M^{EC}_{1,+1\to \frac{1}{2},+\frac{1}{2}} &=&M^{EC}_{1,+1\to 
\frac{1}{2},-\frac{1}{2}} = 
M^{EC}_{1,0\to \frac{1}{2},-\frac{1}{2}} = M^{EC}_{1,-1 \to \frac{1}{2},+\frac{1}{2}} = 0,
\nonumber\\
 \hspace{-0.3in} M^{EC}_{1,0\to \frac{1}{2},+\frac{1}{2}} &=& \sqrt{2 M_m 2E_d  E_{\nu}}
\,\tilde{\cal M}^{EC}_{\rm GT}\,\frac{1}{\sqrt{2}}\,[\varphi^{\dagger}_{n,+\frac{1}{2}}
\vec{\sigma}\,\varphi_{p,+\frac{1}{2}} - \varphi^{\dagger}_{n,-\frac{1}{2}}
\vec{\sigma}\,\varphi_{p,-\frac{1}{2}}]\nonumber\\
\hspace{-0.3in}&&\cdot [\varphi^{\dagger}_{\nu,-\frac{1}{2}}(1 - \vec{n}\cdot\vec{\sigma}\,)
\vec{\sigma}\,\varphi_{e,-\frac{1}{2}}],\nonumber\\
\hspace{-0.3in} M^{EC}_{1,-1 \to \frac{1}{2},-\frac{1}{2}} &=& \sqrt{2 M_m 2E_d E_{\nu}}\,
\tilde{\cal M}^{EC}_{\rm GT}\,
[\varphi^{\dagger}_{n,+\frac{1}{2}}
\vec{\sigma}\,\varphi_{p,-\frac{1}{2}}]\cdot 
[\varphi^{\dagger}_{\nu,-\frac{1}{2}}(1 - \vec{n}\cdot\vec{\sigma}\,)
\vec{\sigma}\,\varphi_{e,+\frac{1}{2}}].
\end{eqnarray}
Having calculated the spinorial matrix elements we obtain
\begin{eqnarray}\label{labelA.28} 
  \hspace{-0.3in} M^{EC}_{1,+1\to \frac{1}{2},+\frac{1}{2}} &=& 
M^{EC}_{1,+1\to \frac{1}{2},-\frac{1}{2}} 
= M^{EC}_{1,0\to \frac{1}{2},-\frac{1}{2}}= M^{EC}_{1,-1 \to \frac{1}{2},+\frac{1}{2}} = 0,
\nonumber\\
  \hspace{-0.3in} M^{EC}_{1,0\to \frac{1}{2},+\frac{1}{2}} &=& -
2\sqrt{2} \sqrt{2 M_m 2E_d E_{\nu}}\,\tilde{\cal M}^{EC}_{\rm GT} =
-\, \sqrt{2 M_m 2E_d E_{\nu}}\,{\cal M}_{\rm GT}\, \langle
\psi^{(Z-1)}_{1s} \psi^{(Z)}_{(1s)^2}\rangle,\nonumber\\
  \hspace{-0.3in} M^{EC}_{1,-1 \to \frac{1}{2},-\frac{1}{2}} &=& 4\,\sqrt{2 M_m 2E_d E_{\nu}}\,
  \tilde{\cal M}^{EC}_{\rm GT} = \sqrt{2}\,\sqrt{2 M_m 2E_d E_{\nu}}\,{\cal M}_{\rm GT}\,
 \langle
\psi^{(Z-1)}_{1s} \psi^{(Z)}_{(1s)^2}\rangle,
\end{eqnarray}
where we have used Eq.(\ref{labelA.26}).  As a result the $EC$--decay
constant of the He--like ${^{140}}{\rm Pr}^{57+}_{I = 1}$ ion is
defined by
\begin{eqnarray}\label{labelA.29} 
  \lambda^{(\rm He)}_{EC} &=& \frac{1}{2M_m}\,\frac{1}{2 I + 1}\int (|M^{EC}_{1,0\to \frac{1}{2},+\frac{1}{2}}|^2  +
  |M^{EC}_{1,-1 \to \frac{1}{2},-\frac{1}{2}}|^2)
  (2\pi)^4\delta^{(4)}(p_d + k - p_m)\nonumber\\
  &&\times\,\frac{d^3p_d}{(2\pi)^3 2E_d}\frac{d^3k}{(2\pi)^3 2E_{\nu}} = \frac{1}{2 I + 1}\,\frac{3}{2}\,
|{\cal M}_{\rm GT}|^2\,|\langle \psi^{(Z-1)}_{1s}\psi^{(Z)}_{(1s)^2}\rangle|^2\,
  \frac{Q^2_{\rm He}}{\pi}.
\end{eqnarray}
Thus, the $EC$--decay constant of the He--like ${^{140}}{\rm
Pr}^{57+}$ ion is equal to
\begin{eqnarray}\label{labelA.30} 
  \lambda^{(\rm He)}_{EC} = \frac{1}{2 I + 1}\,\frac{3}{2}\,|{\cal M}_{\rm GT}|^2\,
|\langle \psi^{(Z-1)}_{1s}\psi^{(Z)}_{(1s)^2}\rangle|^2\,
\frac{Q^2_{\rm He}}{\pi},
\end{eqnarray}
where $Q_{\rm He} = (3352\pm 6)\,{\rm keV}$ is the $Q$--value of the
$EC$--decay of the He--like ${^{140}}{\rm Pr}^{57+}$ ion.

\subsubsection*{$\beta^+$--decay of the He--like ${^{140}}{\rm Pr}^{57+}$ ion}

The $\beta^+$--decay ${^{140}}{\rm Pr}^{57+}_{I = 1}\to {^{140}}{\rm
Ce}^{56+}_{I' = 0} + e^+ + \nu_e$ describes a transition of the
He--like mother ion ${^{140}}{\rm Pr}^{57+}_{I = 1}$ from the ground
$(1s)^2$ state $|I, I_z\rangle$ with $I = 1$ and $I_z = 0,\pm 1$ into
the He--like daughter ion ${^{140}}{\rm Ce}^{56+}_{I' = 0}$ in the
ground $(1s)^2$ state $|I',I'_z\rangle$ with $I' = 0$ and $I'_z = 0$.
The $\beta^+$--decay constant of the He--like ${^{140}}{\rm Pr}^{57+}$
ion is defined by
\begin{eqnarray}\label{labelA.31} 
  \lambda^{(\rm He)}_{\beta^+} &=& \frac{1}{2M_m}\,\frac{1}{2 I + 1}
\sum_{I_z =0,\pm 1}\int |M^{\beta^+}_{_{I,I_z \to 00}}|^2\,
(2\pi)^4\delta^{(4)}(p_d + k + p_+ - p_m)\,F(Z - 1, E_+)\nonumber\\
&&\times\, \frac{d^3p_d}{(2\pi)^3 2E_d}\frac{d^3k}{(2\pi)^3 2E_{\nu}}
\frac{d^3p_+}{(2\pi)^3 2E_+},
\end{eqnarray}
Using the results obtained above and following \cite{APR5,HS66,EK66},
for the amplitudes $M^{\beta^+}_{_{I,I_z \to 00}}$ of the
$\beta^+$--decay ${^{140}}{\rm Pr}^{57+}_{I = 1}\to {^{140}}{\rm
Ce}^{56+}_{I' = 0} + e^+ + \nu_e$ we obtain the following expressions
\begin{eqnarray}\label{labelA.32} 
 \hspace{-0.3in} M^{\beta^+}_{1,+1\to 00} &=& \sqrt{2 M_m 2E_d}\,
\frac{{\cal M}_{\rm GT}}{2\sqrt{2}}\,[\varphi^{\dagger}_{n,-\frac{1}{2}}
\vec{\sigma}\,\varphi_{p,+\frac{1}{2}}]\cdot [\bar{u}_{\nu}(\vec{k},-\frac{1}{2})\vec{\gamma}\,
(1 - \gamma^5) v_{e^+}(\vec{p}_+,\sigma_+)],\nonumber\\
 \hspace{-0.3in} M^{\beta^+}_{1,0\to 00} &=& \sqrt{2 M_m 2E_d }\,\frac{{\cal M}_{\rm GT}}{2\sqrt{2}}
\,\frac{1}{\sqrt{2}}\,[\varphi^{\dagger}_{n,+\frac{1}{2}}
\vec{\sigma}\,\varphi_{p,+\frac{1}{2}} - \varphi^{\dagger}_{n,-\frac{1}{2}}
\vec{\sigma}\,\varphi_{p,-\frac{1}{2}}]\nonumber\\
\hspace{-0.3in}&&\cdot [\bar{u}_{\nu}(\vec{k},-\frac{1}{2})\vec{\gamma}\,(1 - \gamma^5) 
v_{e^+}(\vec{p}_+,\sigma_+)],\nonumber\\
 \hspace{-0.3in} M^{\beta^+}_{1,-1 \to 00} &=& \sqrt{2 M_m 2E_d }\,\frac{{\cal M}_{\rm GT}}{2\sqrt{2}}
[\varphi^{\dagger}_{n,+\frac{1}{2}}
\vec{\sigma}\,\varphi_{p,-\frac{1}{2}}]\cdot [\bar{u}_{\nu}(\vec{k},-\frac{1}{2})\vec{\gamma}\,
(1 - \gamma^5) v_{e^+}(\vec{p}_+,\sigma_+)].\nonumber\\
 \hspace{-0.3in}&&
\end{eqnarray}
Using Eq.(\ref{labelA.16}) for the calculation of the proton--neutron
matrix elements we get
\begin{eqnarray}\label{labelA.33} 
  \hspace{-0.3in} M^{\beta^+}_{1,+1\to 00} &=& \sqrt{2 M_m 2E_d}\,
  \frac{{\cal M}_{\rm GT}}{2\sqrt{2}}\,[\bar{u}_{\nu}(\vec{k},-\frac{1}{2})(\gamma_1 - i\gamma_2)\,
  (1 - \gamma^5) v_{e^+}(\vec{p}_+,\sigma_+)],\nonumber\\
  \hspace{-0.3in} M^{\beta^+}_{1,0\to 00} &=& \sqrt{2 M_m 2E_d }\,\frac{{\cal M}_{\rm GT}}{2\sqrt{2}}
  \,\sqrt{2}\,[\bar{u}_{\nu}(\vec{k},-\frac{1}{2})\gamma_3\,(1 - \gamma^5) 
  v_{e^+}(\vec{p}_+,\sigma_+)],\nonumber\\
  \hspace{-0.3in} M^{\beta^+}_{1,-1 \to 00} &=& \sqrt{2 M_m 2E_d }\,\frac{{\cal M}_{\rm GT}}{2\sqrt{2}}
 [\bar{u}_{\nu}(\vec{k},-\frac{1}{2})(\gamma_1 + i\gamma_2)\,
  (1 - \gamma^5) v_{e^+}(\vec{p}_+,\sigma_+)].
\end{eqnarray}
The squared absolute values of the amplitudes $M^{\beta^+}_{1,+ 1 \to
00}$, $M^{\beta^+}_{1,0\to 00}$ and $M^{\beta^+}_{1,-1 \to 00}$ are
\begin{eqnarray}\label{labelA.34} 
  \hspace{-0.3in}  |M^{\beta^+}_{1,+1 \to 00}|^2 &=& 2 M_m 2E_d\,\frac{|{\cal M}_{\rm GT}|^2}{8}
  {\rm tr}\{\hat{k}(\gamma_1 + i\gamma_2)(1 - \gamma^5)(\hat{p}_+ - m_e)(\gamma_1 - i\gamma_2)
  (1 - \gamma^5)\}=\nonumber\\
  &=& 2 M_m 2E_d\,|{\cal M}_{\rm GT}|^2\,2\,(E_{\nu} E_+ - k_z p_{+z}),\nonumber\\
  \hspace{-0.3in} |M^{\beta^+}_{1,0\to 00}|^2 &=& 2 M_m 2E_d\,\frac{|{\cal M}_{\rm GT}|^2}{8}\,2\,
  {\rm tr}\{\hat{k}\gamma_3(1 - \gamma^5)(\hat{p}_+ - m_e)\gamma_3 (1 - \gamma^5)\} =\nonumber\\
  &=& 2 M_m 2E_d\,|{\cal M}_{\rm GT}|^2\,2\,(E_{\nu} E_+ + k_z p_{+z}),\nonumber\\
  \hspace{-0.3in}  |M^{\beta^+}_{1,-1 \to 00}|^2 &=& 2 M_m 2E_d\,\frac{|{\cal M}_{\rm GT}|^2}{8}\,
  {\rm tr}\{\hat{k}(\gamma_1 - i\gamma_2)(1 - \gamma^5)(\hat{p}_+ - m_e)(\gamma_1 + i\gamma_2)
  (1 - \gamma^5)\}=\nonumber\\
  &=& 2 M_m 2E_d\,|{\cal M}_{\rm GT}|^2\,2\,(E_{\nu} E_+ - k_z p_{+z}),
\end{eqnarray}
where we have used Eq.(\ref{labelA.19}).  The sum of the squared
absolute values of the amplitudes Eq.(\ref{labelA.34}) is
\begin{eqnarray}\label{labelA.35} 
  \sum_{I_z}|M^{\beta^+}_{I,I_z \to 00}|^2 = 2 M_m 2E_d\,|{\cal M}_{\rm GT}|^2\,6\,
(E_{\nu} E_+ - \frac{1}{3}\,k_z p_{+z}).
\end{eqnarray}
Substituting Eq.(\ref{labelA.35}) into Eq.(\ref{labelA.31}) and
integrating over the phase volume of the final state we get
\begin{eqnarray}\label{labelA.36} 
  \lambda^{(\rm He)}_{\beta^+} &=&\frac{3}{2 I + 1}\,|{\cal M}_{\rm GT}|^2
\int \delta^{(4)}(p_d + k + p_+ - p_m)\,F(Z - 1, E_+)\frac{d^3p_dd ^3k d^3p_+}{64\pi^5} = 
\nonumber\\
&=& \frac{3}{2 I + 1}\,\frac{|{\cal M}_{\rm GT}|^2}{4\pi^3}\,f(Q_{\beta^+}, Z - 1),
 \end{eqnarray}
 where $f(Q_{\beta^+}, Z - 1)$ is defined by Eq.(\ref{labelA.22}).
 Thus, the $\beta^+$--decay constant of the He--like ${^{140}}{\rm
 Pr}^{57+}$ ion is equal to
 \begin{eqnarray}\label{labelA.37}
  \lambda^{(\rm He)}_{\beta^+} =  \frac{3}{2 I + 1}\,
\frac{|{\cal M}_{\rm GT}|^2}{4\pi^3}\,f(Q_{\beta^+}, Z - 1),
 \end{eqnarray}
 where $Q_{\beta^+} = (3396\pm 6)\,{\rm keV}$ is the $Q$--value of the
 $\beta^+$--decay of the He--like ${^{140}}{\rm Pr}^{57+}$ ion.

 \section*{Appendix B: Calculation of the $EC$--decay constant of the H-like  
${^{140}}{\rm Pr}^{58+}$ ion from the hyperfine state ${^{140}}{\rm
Pr}^{58+}_{F = \frac{3}{2}}$ }
\renewcommand{\theequation}{B-\arabic{equation}}
\setcounter{equation}{0}

The calculation of the $EC$--decay constant of the H--like
${^{140}}{\rm Pr}^{58+}$ ion from the hyperfine state ${^{140}}{\rm
Pr}^{58+}_{F = \frac{3}{2}}$ with $F = \frac{3}{2}$ is similar to that
for the $EC$--decay from the hyperfine ground state ${^{140}}{\rm
Pr}^{58+}_{F = \frac{1}{2}}$ with $F = \frac{1}{2}$. The $EC$--decay
constant is defined by
\begin{eqnarray}\label{labelB.1} 
 \hspace{-0.3in}&& \lambda^{(\rm H)}_{EC} = \frac{1}{2M_m}\,
\frac{1}{2 F + 1}\sum_{M_F = \pm \frac{1}{2},\pm \frac{3}{2}}
\int |M^{EC}_{_{F,M_F}}|^2\,(2\pi)^4\delta^{(4)}(p_d + k - p_m)\,
\frac{d^3p_d}{(2\pi)^3 2E_d}\frac{d^3k}{(2\pi)^3 2E_{\nu}},\nonumber\\
\hspace{-0.3in}&&
\end{eqnarray}
where $p_m$, $p_d$ and $k$ are 4--momenta of the mother ion, the
daughter ion and the neutrino, respectively. The amplitudes
$M^{EC}_{F,M_F}$ of the $EC$--decay are given by
\begin{eqnarray}\label{labelB.2} 
  \hspace{-0.3in} M^{EC}_{\frac{3}{2},+\frac{3}{2}} &=& 2\,\sqrt{2 M_m 2E_d E_{\nu}}\,
  {\cal M}^{EC}_{\rm GT}\,[\varphi^{\dagger}_{n,-\frac{1}{2}}
  \vec{\sigma}\,\varphi_{p,+\frac{1}{2}}]\cdot 
  [\varphi^{\dagger}_{\nu,-\frac{1}{2}}\,
  \vec{\sigma}\,\varphi_{e,+\frac{1}{2}}],\nonumber\\
  \hspace{-0.3in} M^{EC}_{\frac{3}{2},+\frac{1}{2}} &=& 2\,\sqrt{2 M_m 2E_d E_{\nu}}\,{\cal M}^{EC}_{\rm GT}
  \,\Bigg\{\sqrt{\frac{1}{3}}\,[\varphi^{\dagger}_{n,-\frac{1}{2}}
  \vec{\sigma}\,\varphi_{p,+ \frac{1}{2}}]\cdot 
  [\varphi^{\dagger}_{\nu,-\frac{1}{2}}\,
  \vec{\sigma}\,\varphi_{e,-\frac{1}{2}}]\nonumber\\
  \hspace{-0.3in}&& + \sqrt{\frac{1}{3}}\,[\varphi^{\dagger}_{n,+\frac{1}{2}}
  \vec{\sigma}\,\varphi_{p,+\frac{1}{2}} - \varphi^{\dagger}_{n,-\frac{1}{2}}
  \vec{\sigma}\,\varphi_{p,-\frac{1}{2}}]\cdot 
  [\varphi^{\dagger}_{\nu,-\frac{1}{2}}\,
  \vec{\sigma}\,\varphi_{e,+\frac{1}{2}}]\Bigg\},\nonumber\\
  \hspace{-0.3in} M^{EC}_{\frac{3}{2},-\frac{1}{2}} &=& 2\,\sqrt{2 M_m 2E_d E_{\nu}}\,{\cal M}^{EC}_{\rm GT}
  \,\Bigg\{\sqrt{\frac{1}{3}}\,[\varphi^{\dagger}_{n,+\frac{1}{2}}
  \vec{\sigma}\,\varphi_{p,-\frac{1}{2}}]\cdot 
  [\varphi^{\dagger}_{\nu,-\frac{1}{2}}\,
  \vec{\sigma}\,\varphi_{e,+\frac{1}{2}}]\nonumber\\
  \hspace{-0.3in}&& + \sqrt{\frac{1}{3}}\,[\varphi^{\dagger}_{n,+\frac{1}{2}}
  \vec{\sigma}\,\varphi_{p,+\frac{1}{2}} - \varphi^{\dagger}_{n,-\frac{1}{2}}
  \vec{\sigma}\,\varphi_{p,-\frac{1}{2}}]\cdot 
  [\varphi^{\dagger}_{\nu,-\frac{1}{2}}\,
  \vec{\sigma}\,\varphi_{e,-\frac{1}{2}}]\Bigg\},\nonumber\\
  \hspace{-0.3in} M^{EC}_{\frac{3}{2},-\frac{3}{2}} &=& 2\,\sqrt{2 M_m 2E_d E_{\nu}}\,
  {\cal M}^{EC}_{\rm GT}\,[\varphi^{\dagger}_{n,+ \frac{1}{2}}
  \vec{\sigma}\,\varphi_{p,-\frac{1}{2}}]\cdot 
  [\varphi^{\dagger}_{\nu,-\frac{1}{2}}\,
  \vec{\sigma}\,\varphi_{e,-\frac{1}{2}}].
\end{eqnarray}
Since the spinorial matrix elements vanish
\begin{eqnarray}\label{labelB.3} 
  \hspace{-0.3in} [\varphi^{\dagger}_{n,-\frac{1}{2}}
  \vec{\sigma}\,\varphi_{p,+\frac{1}{2}}]\cdot 
  [\varphi^{\dagger}_{\nu,-\frac{1}{2}}\,
  \vec{\sigma}\,\varphi_{e,+\frac{1}{2}}] = 0,\nonumber\\
  \hspace{-0.3in} [\varphi^{\dagger}_{n,-\frac{1}{2}}
  \vec{\sigma}\,\varphi_{p,+ \frac{1}{2}}]\cdot 
  [\varphi^{\dagger}_{\nu,-\frac{1}{2}}\,
  \vec{\sigma}\,\varphi_{e,-\frac{1}{2}}] = 0,\nonumber\\
\hspace{-0.3in} [\varphi^{\dagger}_{n,+\frac{1}{2}}
  \vec{\sigma}\,\varphi_{p,+\frac{1}{2}} - \varphi^{\dagger}_{n,-\frac{1}{2}}
  \vec{\sigma}\,\varphi_{p,-\frac{1}{2}}]\cdot 
  [\varphi^{\dagger}_{\nu,-\frac{1}{2}}\,
  \vec{\sigma}\,\varphi_{e,+\frac{1}{2}}] = 0,\nonumber\\
  \hspace{-0.3in} [\varphi^{\dagger}_{n,+\frac{1}{2}}
  \vec{\sigma}\,\varphi_{p,-\frac{1}{2}}]\cdot 
  [\varphi^{\dagger}_{\nu,-\frac{1}{2}}\,
  \vec{\sigma}\,\varphi_{e,+\frac{1}{2}}] + [\varphi^{\dagger}_{n,+\frac{1}{2}}
  \vec{\sigma}\,\varphi_{p,+\frac{1}{2}} - \varphi^{\dagger}_{n,-\frac{1}{2}}
  \vec{\sigma}\,\varphi_{p,-\frac{1}{2}}]\cdot 
  [\varphi^{\dagger}_{\nu,-\frac{1}{2}}\,
  \vec{\sigma}\,\varphi_{e,-\frac{1}{2}}] = 0,\nonumber\\
  \hspace{-0.3in}[\varphi^{\dagger}_{n,+ \frac{1}{2}}
  \vec{\sigma}\,\varphi_{p,-\frac{1}{2}}]\cdot 
  [\varphi^{\dagger}_{\nu,-\frac{1}{2}}\,
  \vec{\sigma}\,\varphi_{e,-\frac{1}{2}}] = 0,
\end{eqnarray}
the amplitudes Eq.(\ref{labelB.2}) of the $EC$--decay from the
hyperfine state ${^{140}}{\rm Pr}^{58+}_{F = \frac{3}{2}}$ are equal
to zero
\begin{eqnarray}\label{labelB.4} 
  M^{EC}_{\frac{3}{2},+\frac{3}{2}} = M^{EC}_{\frac{3}{2},+\frac{1}{2}} = 
M^{EC}_{\frac{3}{2},-\frac{1}{2}} = 
  M^{EC}_{\frac{3}{2},-\frac{3}{2}} = 0.
\end{eqnarray}
This confirms the suppression of the $EC$--decay of the H--like
${^{140}}{\rm Pr}^{58+}$ ion from the hyperfine state ${^{140}}{\rm
Pr}^{58+}_{F = \frac{3}{2}}$ pointed out in \cite{GSI,HFS}.


\begin{thebibliography}{9}
\bibitem{He1} $\epsilon^{\rm (He)}_{(1s)^2}(Z) = -\,94.723\,{\rm keV}$
  and $\epsilon^{\rm (He)}_{(1s)^2}(Z - 1) = -\,91.539\,{\rm keV}$ are
  binding energies of the He--like ions ${^{140}}{\rm Pr}^{57+}$
  and ${^{140}}{\rm Ce}^{56+}$ with $Z = 59$.
\end{thebibliography}
\end{document}